\documentclass[12pt,draftcls,onecolumn]{IEEEtran}
\usepackage{cite}
\usepackage[font=small]{caption}
\usepackage{ifpdf}
\usepackage{authblk}
\usepackage{commath}
\usepackage{cuted} 
\usepackage{pgfplots}
\usetikzlibrary{arrows,decorations.markings}
\usepackage{bigints}
\newif\ifCLASSOPTIONonecolumn       \CLASSOPTIONonecolumnfalse
\newif\ifCLASSOPTIONtwocolumn       \CLASSOPTIONtwocolumntrue
\usepackage{tikz}
\ifCLASSINFOpdf

\usepackage{subcaption}
\usepackage{amsmath}
\usepackage{dsfont}
\usepackage{bbold}
\usepackage{amssymb}
\usepackage{algorithmic}
\usepackage{array}
\usepackage{stix}
\usepackage{stfloats}
\fnbelowfloat
\usepackage{bigints}
\begin{document}
%
\title{On the Outage Probability of Vehicular Communications at Intersections Over Nakagami-$m$ Fading Channels}
%
%
%

\author[1]{Baha Eddine Youcef~Belmekki}
\author[2]{Abdelkrim ~Hamza}
\author[1]{Beno\^it~Escrig}
\affil[1]{IRIT Laboratory, School
of ENSEEIHT, Institut National Polytechnique de Toulouse, France,}
\affil[ ]{ e-mail: $\{$bahaeddine.belmekki, benoit.escrig$\}$@enseeiht.fr}
\affil[2]{LISIC Laboratory, Electronic and Computer Faculty, USTHB, Algiers, Algeria,}
\affil[ ]{email: ahamza@usthb.dz}
\setcounter{Maxaffil}{0}
\renewcommand\Affilfont{\small}

\markboth{ }%
{Shell \MakeLowercase{\textit{et al.}}: Bare Demo of IEEEtran.cls for IEEE Journals}

\maketitle

\IEEEpeerreviewmaketitle
 \begin{abstract}
In this paper, we study  vehicular communications (VCs) at intersections, in the presence of interference over Nakagami-$m$ fading channels. The interference are originated from vehicles located on two perpendicular roads. We derive the outage probability, and closed forms are obtained. The outage probability is derived when the destination is on the road (vehicle, cyclist, pedestrian) or outside the road (base station, road side unit). We compare the performance of line of sight (LOS) scenarios and non-line of sight (NLOS) scenarios, and show that NLOS scenarios offer better performance than LOS scenarios. We also compare intersection scenarios with highway scenarios, and show that the performance of intersection scenarios are worst than highway scenarios as the destination moves toward the intersection. Finally, we investigate the performance of VCs in a realistic scenario involving several lanes. All the analytical results are validated by Monte-Carlo simulations.
\end{abstract}


\section{Introduction}
\subsection{Motivation}
Road traffic safety is a major issue, and more particularly at intersections \cite{traficsafety}. Vehicular communications (VCs) offer several applications regarding accident prevention, such as sending safety messages that alert vehicles about accidents happening in their surrounding. One of the major drawbacks that effect VCs are interference. Hence, investigating the performance of VCs in the presence of interference is crucial in order to design safety applications at urban and suburban intersections. 

\subsection{Related Works}
The impact of the presence of interference in VCs considering highway scenarios has been investigated in \cite{blaszczyszyn2009performance,blaszczyszyn2013stochastic,blaszczyszyn2012vehicular}. In \cite{farooq2016stochastic}, the authors derive the expressions for the intensity of concurrent transmitters and packet success probability in multi-lane highways with carrier sense multiple access (CSMA) protocols. The performance of IEEE 802.11p using tools from queuing theory and stochastic geometry is analyzed in \cite{tong2016stochastic}. The authors in \cite{tassi2017modeling} derivate the outage probability and rate coverage probability of vehicles, when the line of sight (LOS) path to the base station is obstructed by large vehicles sharing other highway lanes. In \cite{al2018stochastic}, the performance of automotive radar is evaluated in terms of expected signal-to-noise ratio, when the locations of vehicles follow a Poisson point process and a Bernoulli lattice process.\\
However, few works studied the effect of interference in VCs at intersections. Steinmetz et al derivate the success probability when the received node and the interferer nodes are aligned on the road \cite{steinmetz2015stochastic}.  In \cite{abdulla2016vehicle}, the authors analyze the performance in terms of success probability for finite road segments under different channel conditions. The authors in \cite{abdulla2017fine} evaluate the average and the fine-grained reliability for interference-limited vehicle-to-vehicle (V2V) communications with the use of the meta distribution. In \cite{jeyaraj2017reliability}, the authors analyze the performance of an orthogonal street system which consists of multiple intersections, and show that, in high-reliability regime, the orthogonal street system behaves like a one dimensional Poisson network. However, in low-reliability regime, it behaves like a two dimensional Poisson network. The authors in \cite{kimura2017theoretical} derive the outage probability of V2V communications at intersections in the presence of interference with a power control strategy.

The authors of this the paper investigated the impact of NOMA using direct transmission in \cite{C1,J3},  cooperative NOMA at intersections in \cite{belmekki2019outage,J4}, and MRC using NOMA \cite{Belmkki}, and in millimeter wave vehicular communications in \cite{Cmm1,Cmm2}. The authors of this paper also investigated the impact of vehicles mobility, and different transmission schemes on the performance in \cite{J2} and \cite{J1}, respectively.

Following this line of research, we study the performance of VCs at urban and suburban intersections in the presence of interference.  

\subsection{Contributions}
The  contributions of this paper are  as follows:
\begin{itemize}
\item We derived the outage probability expressions over Nakagami-$m$ fading channels for specific channel conditions, and when the destination is either on the road (V2V) or outside the road (V2I). 
\item
We compare the performance of suburban (LOS) environment and urban (NLOS) environment, and show that, the urban environment  exhibits better performance than suburban environment.

\item We compare intersection scenarios with highway scenarios, and show that the performance of intersection scenarios are worst than highway scenarios as the destination moves toward the intersection. 
\item We investigate a realistic scenario involving several lanes, and show that the increase of lanes decreases the performance.
\end{itemize}



\section{System Model}

\begin{figure}[]
\centering
\includegraphics[scale=0.7]{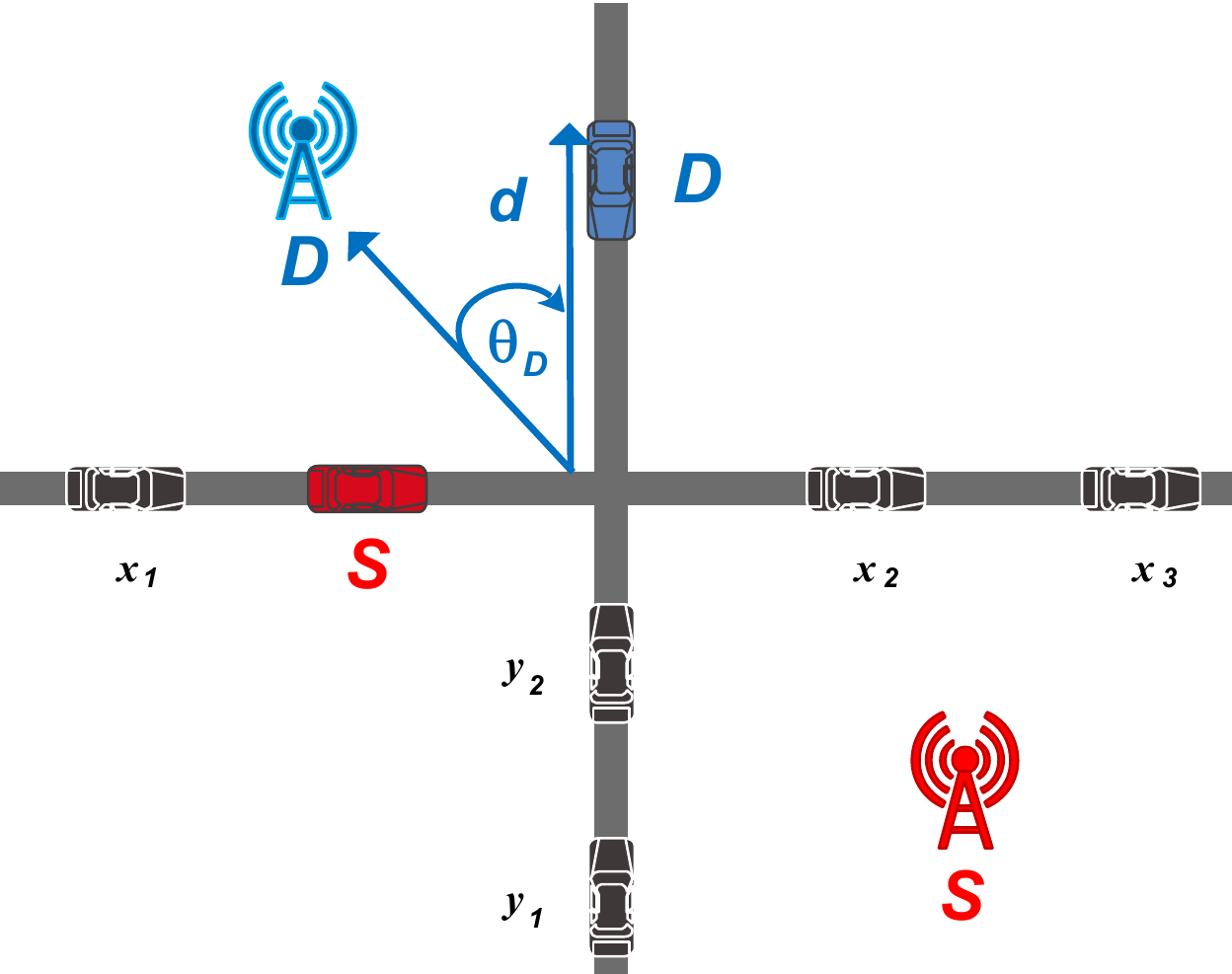}
\caption{System model for VCs at the intersection. The nodes $S$ and $D$ can either be vehicles or as part of communication infrastructures.}
\label{Fig1}
\end{figure}

In this paper, we consider a transmission between a source, denoted $S$, and a destination, denoted $D$. We denote by $S$ and $D$ the nodes and their locations.
We consider an intersection scenario with two perpendicular roads, an horizontal road denoted $X$, and a vertical road denoted $Y$.
In this paper, we consider both V2V and V2I communications, hence, $S$ and $D$ can be either on the road (e.g., vehicle) or outside the road (e.g., base station). We denote by $d$ the distance between $D$ and the intersection as shown in Fig.\ref{Fig1}. Note that the intersection is the point where the $X$ road and the $Y$ road intersect.

The transmission is subject to interference originated from vehicles located on the roads. The set of interfering vehicles located on the $X$ road, denoted by $\Phi_{X}$ (resp. on the $Y$ road, denoted by $\Phi_{Y}$) are modeled as a One-Dimensional Homogeneous Poisson Point Process (1D-HPPP), that is, $\Phi_{X}\sim\textrm{1D-HPPP}(\lambda_{X},x)$ (resp.$\Phi_{Y}$ $\sim \textrm{1D-HPPP}(\lambda_{Y},y)$), where $x$ and $\lambda_{X}$ (resp. $y$ and $\lambda_{Y}$) are the position of interferer vehicles and their intensity on the $X$ road (resp. $Y$ road). The notation $x$ and $y$ denotes both the interfering vehicles and their locations.

The medium access protocols used in VCs are mainly based on CSMA schemes.
However, the mathematical derivation considering these protocols might not be possible in our scenario, and closed form expressions are hard to obtain. In addition, \cite{subramanian2012congestion,nguyen2013performance} showed that the performance of CSMA tends to the performance of ALOHA in dense networks. Hence, we assume that vehicles use slotted Aloha MAC protocol with  parameter $p$, i.e., every node can access the medium with a probability $p$

The transmission between the nodes $a$ and $b$ experiences a path loss denoted by $l_{ab}$, where $l_{ab}= \Vert a- b\Vert^{-\alpha}$, and $\alpha$ is the path loss exponent.

We consider an interference limited scenario, hence, the power of noise is set to zero ($\sigma^2=0$). Without loss of generality, we assume  that all nodes transmit with a unit power. 
The signal received at $D$ is expressed as
 \begin{equation}\label{eq.1}
   \mathcal{Y}_{D}=h_{SD}\sqrt{l_{SD}}\:\chi_{S}+ 
 \sum_{x\in \Phi_{X}}h_{Dx}\sqrt{l_{Dx}}\:\chi_{x} 
 +\sum_{y\in \Phi_{Y}}h_{Diy}\sqrt{l_{Dy}}\:\chi_{y}, \nonumber
 \end{equation}
where $\mathcal{Y}_{D}$ is the signal received by $D$.
The messages transmitted by the interfere node $x$ and $y$, are denoted respectively by $ \chi_x$ and $\chi_y $, $h_{SD}$ denotes the fading coefficient between $S$ and $D$, and it is modeled as Nakagami-$m$ fading with parameter $m$ \cite{hu2013maximum}. Therefore, the power fading coefficient between the node $S$ and $D$, denoted $|h_{SD}|^2$, follows a gamma distribution distribution.
The aggregate interference is defined as 
\begin{eqnarray}\label{eq.2}
I_{X}=\sum_{x\in \Phi_{X}}\vert h_{Dx}\vert^{2}l_{Dx}   \\ 
I_{Y}=\sum_{y\in \Phi_{Y}}\vert h_{Dy}\vert^{2}l_{Dy} , 
\end{eqnarray}
where $I_{X} $ denotes the aggregate interference from the $X$ road, $I_{Y}$ denotes the aggregate interference from the $Y$, $\Phi_{X}$ denotes the set of the interferers from the $X$ road, and $\Phi_{Y}$ denotes the set of the interferers from the $Y$ road.
The coefficients $h_{Dx}$ and  $h_{Dy}$ denote the fading between $D$ and $x$, and between $D$ and $y$ respectively. They are modeled as Rayleigh fading \cite{cheng2007mobile}. Then, the power fading coefficients $|h_{Dx}|^2$ and $|h_{Dy}|^2$, follow an exponential distribution with unit mean.

We consider two scenarios, the LOS scenario, and the non-line of sight scenario (NLOS). The LOS scenario models the suburban environment, whereas the NLOS scenario models the urban scenario. We denote by $\alpha_{\textrm{LOS}}$ and $m_{\textrm{LOS}}$, the path exponent and the fading parameter for LOS, and we denote by $\alpha_{\textrm{NLOS}}$ and $m_{\textrm{NLOS}}$ the path exponent and the fading parameter for NLOS. Hence, we have that $m \in \{m_{\textrm{LOS}},m_{\textrm{NLOS}}\}$ $ \alpha \in \{\alpha_{\textrm{LOS}},\alpha_{\textrm{NLOS}}\}$. 

\section{Outage Expressions}
In this section, we calculate the outage probability of the transmission between $S$ and $D$. An outage event occurs when the SIR at $D$ is below a given threshold. 
The SIR at $D$ is defined as
\begin{equation}\label{eq.3}
\textrm{SIR} \triangleq  \dfrac{\vert h_{SD}\vert^{2}l_{SD}}{I_{X}+I_{Y}}.
\end{equation}
The outage event at $D$ is defined as 
\begin{equation}\label{eq.4}
O \triangleq \big[ \textrm{SIR} < \Theta \big],
\end{equation}
where $\Theta$ is the decoding threshold.
The outage probability expression is given as 
\begin{equation}\label{eq.5}
\mathbb{P}(O) = \mathbb{P}\big( \textrm{SIR} < \Theta \big)
=1-\mathbb{P}\big( \textrm{SIR} \geq \Theta \big).
\end{equation}
To calculate $\mathbb{P}\big( \textrm{SIR} \geq \Theta \big)$, we proceed as follows
\begin{align}\label{eq.6}
\mathbb{P}\big( \textrm{SIR} \geq \Theta \big) &=\mathbb{E}_{I_{X},I_Y}\Bigg[\mathbb{P}\Bigg\lbrace\dfrac{\vert h_{SD}\vert^{2}l_{SD}}{I_{X}+I_{Y}} \ge \Theta \Bigg\rbrace\Bigg]\nonumber\\
&=\mathbb{E}_{I_{X},I_Y}\Bigg[\mathbb{P}\Bigg\lbrace\vert h_{SD}\vert^{2}  \ge \dfrac{\Theta}{l_{SD}} \big(I_{X}+I_{Y}\big) \Bigg\rbrace\Bigg].
\end{align}
Since $\vert h_{SD}\vert^{2}$ follows a gamma distribution, its complementary cumulative distribution function (CCDF) is given by
\begin{equation}\label{eq.7}
\bar{F}_{\vert h_{SD}\vert^{2}}(X)=\mathbb{P}(\vert h_{SD}\vert^{2}>X)=\frac{\Gamma(m,\frac{m}{\mu }X)}{\Gamma(m)},
\end{equation}
hence
\begin{equation}\label{eq.7}
\mathbb{P}\big( \textrm{SIR} \geq \Theta \big)=\mathbb{E}_{I_{X},I_Y}\Bigg[\frac{\Gamma\Big(m,\dfrac{m \:\Theta}{\mu\: l_{SD}} (I_{X}+I_{Y})\Big)}{\Gamma(m)}\Bigg].
\end{equation}
The exponential sum function when $m$ is an integer is defined as 
\begin{equation}\label{eq.8}
e_{(m)}=\sum_{k=0}^{m-1} \frac{(\frac{m}{\mu }X)^k}{k!}=e^x \frac{\Gamma(m,\frac{m}{\mu }X)}{\Gamma(m)},
\end{equation}
then
\begin{equation}\label{eq.9}
\frac{\Gamma(m,\frac{m}{\mu }X)}{\Gamma(m)}=e^{-\frac{m}{\mu}X}\sum_{k=0}^{m-1}\frac{1}{k!}{\big(\frac{m\:X}{\mu}\big)}^k.
\end{equation}
The equation (\ref{eq.7}) then becomes
\begin{equation}\label{eq.9.1}
\mathbb{P}\big( \textrm{SIR} \geq \Theta \big)
=\sum_{k=0}^{m-1}\frac{1}{k!}\:\Big(\dfrac{m\:\Theta}{\mu \:l_{SD}}\Big)^k \mathbb{E}_{I_{X},I_Y}\Bigg[\exp\Big(-\dfrac{m \:\Theta}{\mu\: l_{SD}} \big(I_{X}+I_{Y}\big)\Big)\big(I_{X}+I_{Y}\big)^k\Bigg].
\end{equation}
Applying the binomial theorem in (\ref{eq.9.1}), we get
\begin{equation}\label{eq.10}
\mathbb{P}\big( \textrm{SIR} \geq \Theta \big)
=\sum_{k=0}^{m-1}\frac{1}{k!}\:G^k \mathbb{E}_{I_{X},I_Y}\Bigg[\exp\Big(-G \big[I_{X}+I_{Y}\big]\Big)\sum_{n=0}^{k} \binom{k}{n}{I_{X}}^{k-n}\:{I_{Y}}^n\Bigg],
\end{equation}
where $G=\dfrac{m\:\Theta}{\mu \:l_{SD}}$. We denote, by $\mathcal{A}(I_{X},I_Y)$, the expectation in (\ref{eq.10}), hence $\mathcal{A}(I_{X},I_Y)$ becomes
\begin{align}\label{eq.14}
\mathcal{A}(I_{X},I_Y)=&\mathbb{E}_{I_{X},I_Y}\Bigg[e^{-G\:I_{X}}\:e^{-G\:I_{Y}}\sum_{n=0}^{k} \binom{k}{n}{I_{X}}^{k-n}\:{I_{Y}}^n\Bigg]\nonumber \\
=&\sum_{n=0}^{k}\binom{k}{n}\mathbb{E}_{I_{X},I_Y}\Bigg[e^{-G\:I_{X_{D}}}\:e^{-G\:I_{Y}}{I_{X}}^{k-n}\:{I_{Y}}^n\Bigg]\nonumber \\
=&\sum_{n=0}^{k}\binom{k}{n}\mathbb{E}_{I_{X}}\Bigg[e^{-G\:I_{X}}{I_{X}}^{k-n}\Bigg]\mathbb{E}_{I_Y}\Bigg[e^{-G\:I_{Y}}\:{I_{Y}}^n\Bigg]\nonumber \\ 
=&\sum_{n=0}^{k}\binom{k}{n}(-1)^{k-n}\frac{\textrm{d}^{k-n} \mathcal{L}_{I_{X}}(G)}{\textrm{d}^{k-n} G} (-1)^{n}\frac{\textrm{d}^{n} \mathcal{L}_{I_{Y}}(G)}{\textrm{d}^{n} G}\nonumber\\ 
=&(-1)^{k}\sum_{n=0}^{k}\binom{k}{n}\frac{\textrm{d}^{k-n} \mathcal{L}_{I_{X}}(G)}{\textrm{d}^{k-n} G} \frac{\textrm{d}^{n} \mathcal{L}_{I_{Y}}(G)}{\textrm{d}^{n} G}.
\end{align}
Using the following property 
\begin{equation}\label{eq.15}
\mathbb{E}_{I}\Big[e^{-gI}{I}^{N}\Big]=(-1)^N \frac{\textrm{d}^{N}\mathbb{E}_{I}\Big[e^{-gI}{I}^{N}\Big]}{\textrm{d}^{N} g}=(-1)^N\frac{\textrm{d}^{N} \mathcal{L}_{I}(g)}{\textrm{d}^{N} g},
\end{equation}
then 
\begin{equation}\label{eq.16}
\mathbb{P}\big( \textrm{SIR} \geq \Theta \big)=\sum_{k=0}^{m-1}\frac{1}{k!}\:\big(-\dfrac{m\:\Theta}{\mu \:l_{SD}}\big)^k \sum_{n=0}^{k}\binom{k}{n}\frac{\textrm{d}^{k-n} \mathcal{L}_{I_{X}}\big(\dfrac{m\:\Theta}{\mu \:l_{SD}}\big)}{\textrm{d}^{k-n} \big(\dfrac{m\:\Theta}{\mu \:l_{SD}}\big)} \frac{\textrm{d}^{n} \mathcal{L}_{I_{Y}}\big(\dfrac{m\:\Theta}{\mu \:l_{SD}}\big)}{\textrm{d}^{n} \big(\dfrac{m\:\Theta}{\mu \:l_{SD}}\big)}.
\end{equation}
Plugging (\ref{eq.16}) into (\ref{eq.5}) yields of the outage probability expression.
The expression of $\textrm{d}^{k-n} \mathcal{L}_{I_{X}}(s)/\textrm{d}^{k-n} (s)$ and $\textrm{d}^{n} \mathcal{L}_{I_{Y}}(s)/\textrm{d}^{n} (s)$ are given respectively by

 \begin{align}\label{eq.21}
  \frac{\textrm{d}^{n} \mathcal{L}_{I_{X_{D}}}\big(s\big)}{\textrm{d}^{n} s}=& \Bigg[-\frac{1}{4}\frac{\emph{p}\lambda_{Y}\pi(\sqrt{d_y^4+s}-d_y^2)}{\sqrt{2\sqrt{d_y^4+s}+2d_y^2} (d_y^4+s)}\nonumber\\
  &-\frac{1}{4}\frac{\emph{p}\lambda_{Y}\pi\sqrt{2\sqrt{d_y^4+s}+2d_y^2}}{d_y^4+s}+\frac{1}{4}\frac{\emph{p}\lambda_{Y}\pi\sqrt{2\sqrt{d_y^4+s}+2d_y^2}(\sqrt{d_y^4+s}-d_y^2)}{(d_y^4+s)^{3/2}}\Bigg]^n\nonumber\\
&\times\exp\Bigg(-\frac{1}{2}\frac{\emph{p}\lambda_{Y}\pi\sqrt{2\sqrt{d_y^4+s}+2d_y^2}(\sqrt{d_y^4+s}-d_y^2)}{\sqrt{d_y^4+s}}\Bigg),
\end{align}
and
\begin{align}\label{eq.22}
  \frac{\textrm{d}^{n} \mathcal{L}_{I_{Y_{D}}}\big(s\big)}{\textrm{d}^{n} s}=& \Bigg[-\frac{1}{4}\frac{\emph{p}\lambda_{Y}\pi(\sqrt{d_x^4+s}-d_x^2)}{\sqrt{2\sqrt{d_x^4+s}+2d_x^2} (d_x^4+s)}\nonumber\\
  &-\frac{1}{4}\frac{\emph{p}\lambda_{Y}\pi\sqrt{2\sqrt{d_x^4+s}+2d_x^2}}{d_x^4+s}
  +\frac{1}{4}\frac{\emph{p}\lambda_{Y}\pi\sqrt{2\sqrt{d_x^4+s}+2d_x^2}(\sqrt{d_x^4+s}-d_x^2)}{(d_x^4+s)^{3/2}}\Bigg]^n\nonumber\\
&\times\exp\Bigg(-\frac{1}{2}\frac{\emph{p}\lambda_{Y}\pi\sqrt{2\sqrt{d_x^4+s}+2d_x^2}(\sqrt{d_x^4+s}-d_x^2)}{\sqrt{d_x^4+s}}\Bigg).
\end{align}
 \section{Laplace Transform Expressions} 
After we obtained the expression of the outage probability, we derive, in this section, the Laplace transform expressions of the interference from the $X$ road and from the $Y$ road.
The Laplace transform of the interference originating from the $X$ road, is expressed as
\begin{equation}\label{eq.17}
\mathcal{L}_{{I_{X}}}(s)=\mathbb{E}\big[{\\\exp(-sI_{X})}\big].
\end{equation}
Plugging (\ref{eq.2}) into (\ref{eq.17}) yields
\begin{align}\label{eq.18}
\mathcal{L}_{{I_{X}}}(s)=&\mathbb{E}\Bigg[{\exp\Bigg(-\sum_{x\in\Phi_{X}}s\vert h_{D_{x}}\vert^2 l_{D_{x}} \Bigg)}\Bigg]\nonumber\\
=& \mathbb{E}\Bigg[\prod_{x\in\Phi_{X}} \exp\Bigg(-s\vert h_{D_{x}}\vert^2l_{D_{x}}\Bigg)\Bigg]\nonumber\\
\overset{(a)}{=}&\mathbb{E}\Bigg[\prod_{x\in\Phi_{X}}\mathbb{E}_{\vert  h_{D_{x}}\vert^2, p}\Bigg\lbrace \exp\Bigg(-s\vert h_{D_{x}}\vert^2l_{D_{x}}\Bigg)\Bigg\rbrace\Bigg]\nonumber\\
\overset{(b)}{=}&\mathbb{E}\Bigg[\prod_{x\in\Phi_{X}}\dfrac{p}{1+s l_{D_{x}}}+1-p\Bigg] \nonumber\\
\overset{(c)}{=}&\exp\Bigg(-\lambda_{X}\displaystyle\int_{\mathbb{R}}\Bigg[1-\bigg(\dfrac{p}{1+sl_{D_{x}}}+1-p\bigg)\Bigg]\textrm{d}x\Bigg)\nonumber\\
=&\exp\Bigg(-p\lambda_{X}\displaystyle\int_{\mathbb{R}}\dfrac{1}{1+1/sl_{D_{x}}}\textrm{d}x\Bigg), 
\end{align}

where (a) follows from the independence of the fading coefficients; (b) follows from performing the expectation over $|h_{{D}x}|^2$ which follows an exponential distribution with unit mean, and performing the expectation over the set of interferes; (c) follows from the probability generating functional (PGFL) of a PPP \cite{haenggi2012stochastic}. 
Then, substituting $l_{{D}x}=\Vert \textit{x}-D \Vert^{-\alpha}$ in (\ref{eq.18}) yields 
\begin{equation}\label{eq.23}
\mathcal{L}_{I_{X}}(s)=\exp\Bigg(-\emph{p}\lambda_{X}\int_\mathbb{R}\dfrac{1}{1+\Vert \textit{x}-{D} \Vert^\alpha/s}\textrm{d}x\Bigg),
\end{equation}
where 
\begin{equation}\label{eq.24}
\Vert \textit{x}-{D} \Vert=\sqrt{\big[d\sin(\theta_{{D}})\big]^2+\big[x-d \cos(\theta_{D}) \big]^2 }. 
\end{equation}
The Laplace transform of the interference originating from the $Y$ road can be acquired by following the same steps above, and it is given by
\begin{equation}\label{eq.25}
\mathcal{L}_{I_{Y}}(s)=\exp\Bigg(-\emph{p}\lambda_{Y}\int_\mathbb{R}\dfrac{1}{1+\Vert \textit{y}-{D} \Vert^\alpha/s}\textrm{d}y\Bigg),
\end{equation}
where
\begin{equation}\label{eq.26}
\Vert \textit{y}-{D} \Vert=\sqrt{\big[d\cos(\theta_{{D}})\big]^2+\big[y-d \sin(\theta_{D}) \big]^2 },
\end{equation}
where $\theta_{D} $ is the angle between the node ${D}$ and the $X$ road.\\
The expression (\ref{eq.23}) and (\ref{eq.25}) can be calculated with mathematical tools such as MATLAB. Closed form expressions are obtained for $\alpha=2$ and $\alpha=4$. We only present the expressions when $\alpha=4$ due to lack of space.\\
In order to calculate the Laplace transform of interference originated from the X road at $D$, we have to calculate the integral in (\ref{eq.23}) for $\alpha=4$. Let us take $d_{x}=d \cos(\theta_{D})$, and $d_{y}=d \sin(\theta_{D}$), then (\ref{eq.23}) becomes
\begin{multline}\label{eq.27}
\mathcal{L}_{I_{X_{D}}}(s)=\\ \exp\Bigg(-\emph{p}\lambda_{X}\int_\mathbb{R}\dfrac{1}{1+\big({\sqrt{d_{y}^2+(x-d_{x})^2}}\big)^4/s }\textrm{d}x\Bigg)\\
=\exp\Bigg(-\emph{p}\lambda_{X}s\int_\mathbb{R}\dfrac{1}{s+\big({\sqrt{d_{y}^2+(x-d_{x})^2}}\big)^4 }\textrm{d}x\Bigg),
\end{multline}
and the integral inside the exponential in (\ref{eq.27}) equals
\begin{equation}\label{eq.28}
\int_\mathbb{R}\dfrac{1}{s+\big({\sqrt{d_{y}^2+(x-d_{x})^2}}\big)^4}\textrm{d}x=\frac{\sqrt{2\sqrt{{d_y}^4+s}+2{d_y}^2} \: \big(\sqrt{{d_y}^4+s}-{d_y}^2\big)}{2\sqrt{{d_y}^4+s}}.
\end{equation}
Then, plugging (\ref{eq.28}) into (\ref{eq.27}), and substituting $d_{y}$ by $d\sin(\theta_{{D}})$ yields 
\begin{equation}\label{eq.29}
\mathcal{L}_{I_{X}}(s)=\exp\Bigg(-\emph{p}\lambda_{X}\pi\mathcal{V}_x(s)\Bigg),
\end{equation}
where
\begin{equation}\label{eq.30}
\mathcal{V}_x(s)=\frac{\sqrt{2\sqrt{{(d \sin(\theta_{D})}^4+s}+2{(d \sin(\theta_{D})}^2} }{2\sqrt{{(d \sin(\theta_{D})}^4+s}} \times  \sqrt{{(d \sin(\theta_{D})}^4+s}-{(d \sin(\theta_{D})}^2.
\end{equation}
Following the same steps above, and without details for the derivation, the Laplace transform expressions of the interference for an intersection scenario, when $\alpha=4$ are given by
\begin{equation}\label{eq.31}
\mathcal{L}_{I_{Y}}(s)=\exp\Bigg(-\emph{p}\lambda_{Y}\pi\mathcal{V}_y(s)\Bigg),
\end{equation}
where
\begin{equation}\label{eq.32}
\mathcal{V}_y(s)=\frac{\sqrt{2\sqrt{{(d \cos(\theta_{D})}^4+s}+2{(d \cos(\theta_{D})}^2} \: }{2\sqrt{{(d \cos(\theta_{D})}^4+s}}\times \sqrt{{(d \cos(\theta_{D})}^4+s}-{(d \cos(\theta_{D})}^2.
\end{equation}

\section{Simulations and Discussions}
In this section, we evaluate the performance of a transmission between $S$ and $D$ at intersections. In order to verify the accuracy of the theoretical results, Monte Carlo simulations are obtained by averaging over 50,000 realizations of the PPPs and fading parameters. We set our simulation area to $[-10^3 \textrm{m}, 10^3 \textrm{m}]$ for each road. Without loss of generality, we set $\lambda_X = \lambda_Y = \lambda$. Finally, we consider $\alpha_{\textrm{LOS}}=2$, $m_{\textrm{LOS}}=3$, $\alpha_{\textrm{NLOS}}=4$,   $m_{\textrm{NLOS}}=1$, and $\mu=1$.
\begin{figure*}[t!]
    \centering
    \begin{subfigure}[b]{0.5\textwidth}
        \centering
        \includegraphics[height=8cm,width=9cm]{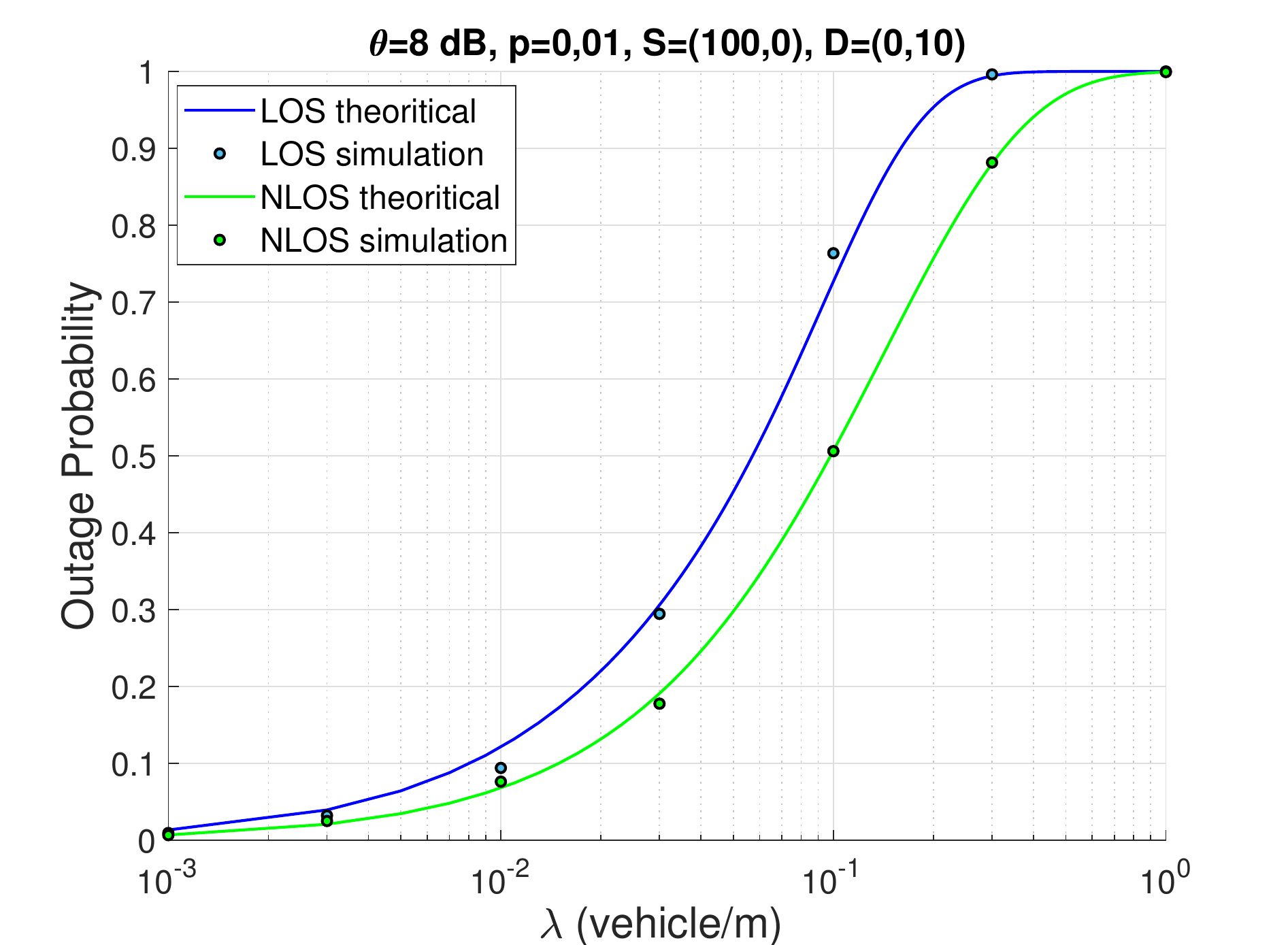}
        \caption{}
        \label{Fig2(a)}
    \end{subfigure}%
    ~ 
    \begin{subfigure}[b]{0.5\textwidth}
        \centering
        \includegraphics[height=8cm,width=9cm]{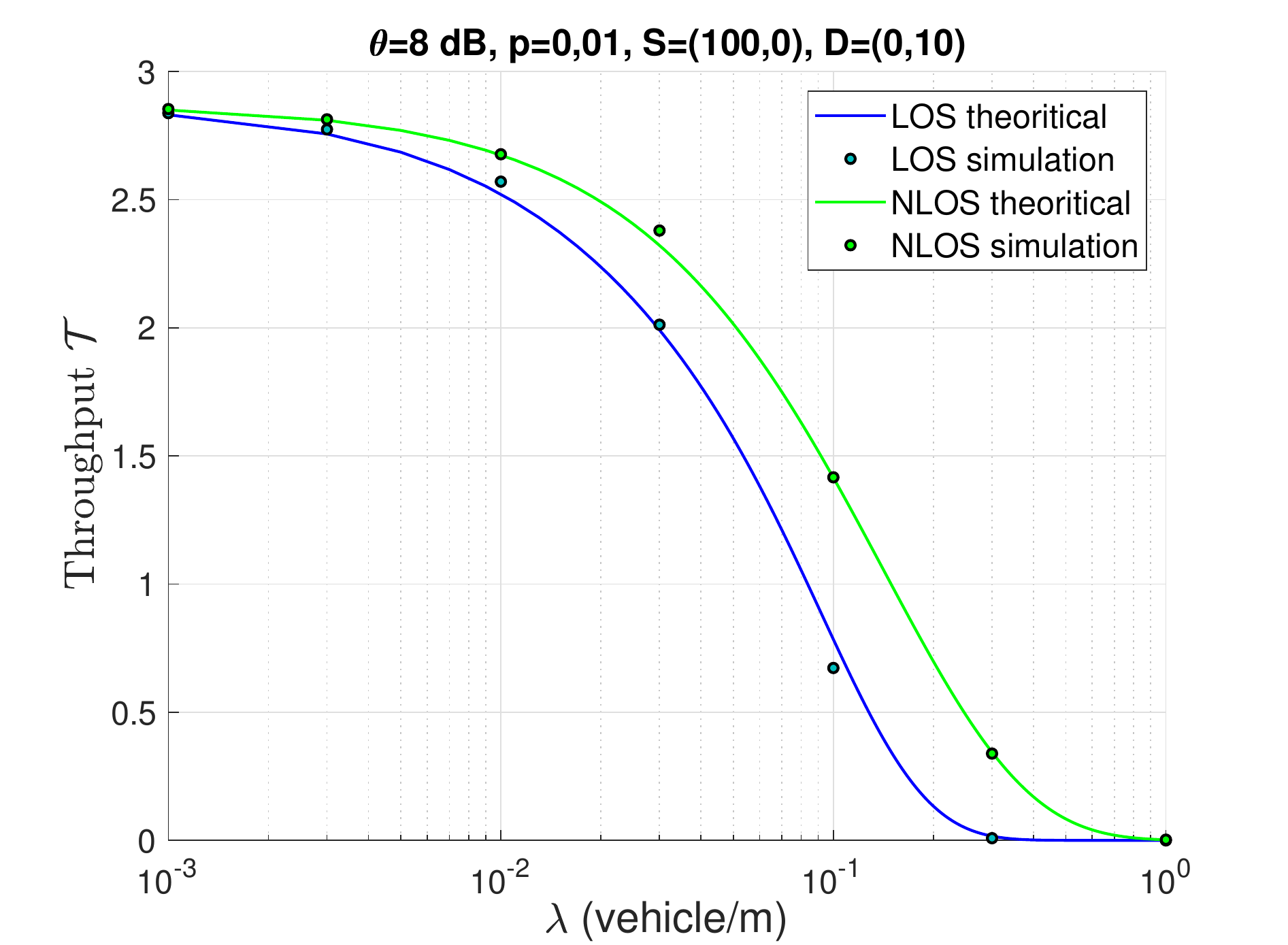}
        \caption{}
        \label{Fig2(b)}
    \end{subfigure}
    \caption{Performance of VCs as a function of the vehicles density $\lambda$. (a) Outage probability for suburban environment (LOS), and urban environment (NLOS). (b)  throughput for suburban environment (LOS), and urban environment (NLOS).}
 \label{Fig2}
\end{figure*}

Fig.\ref{Fig2} depicts the performance of VCs in terms of outage probability and throughput as a function of vehicles density for LOS and NLOS. We define the throughput, denoted $\mathcal{T}$ as follows
\begin{equation}
\mathcal{T}=\mathbb{P}\big( \textrm{SIR} \geq \Theta \big)\times \log_2(1+\Theta). \nonumber
\end{equation} 
We see from Fig.\ref{Fig2} that, as the vehicles intensity increases, the outage probability increases, and the throughput decreases. This is because, as intensity increases, the aggregate interference at the receiver increases as well. Hence, reducing the SIR, which will cause an increase in the outage probability, and a decrease in throughput.  Surprisingly, we can see that, NLOS scenario has a better performance than LOS scenario, even though the fading degree of NLOS is higher than LOS ($m_{\textrm{LOS}} > m_{\textrm{NLOS}}$). This is due to the fact that, $\alpha_{\textrm{NLOS}}>\alpha_{\textrm{LOS}}$. Hence, the signals of interfering vehicles for NLOS scenario decrease rapidly compared to the signals of LOS scenario. Therefore, the aggregate of the interference in LOS scenario is greater than NLOS scenario. 
 We can draw the conclusion that suburban intersections exhibit lower performance that urban intersections.

\begin{figure*}[t!]
    \centering
    \begin{subfigure}[b]{0.5\textwidth}
        \centering
        \includegraphics[height=8cm,width=9cm]{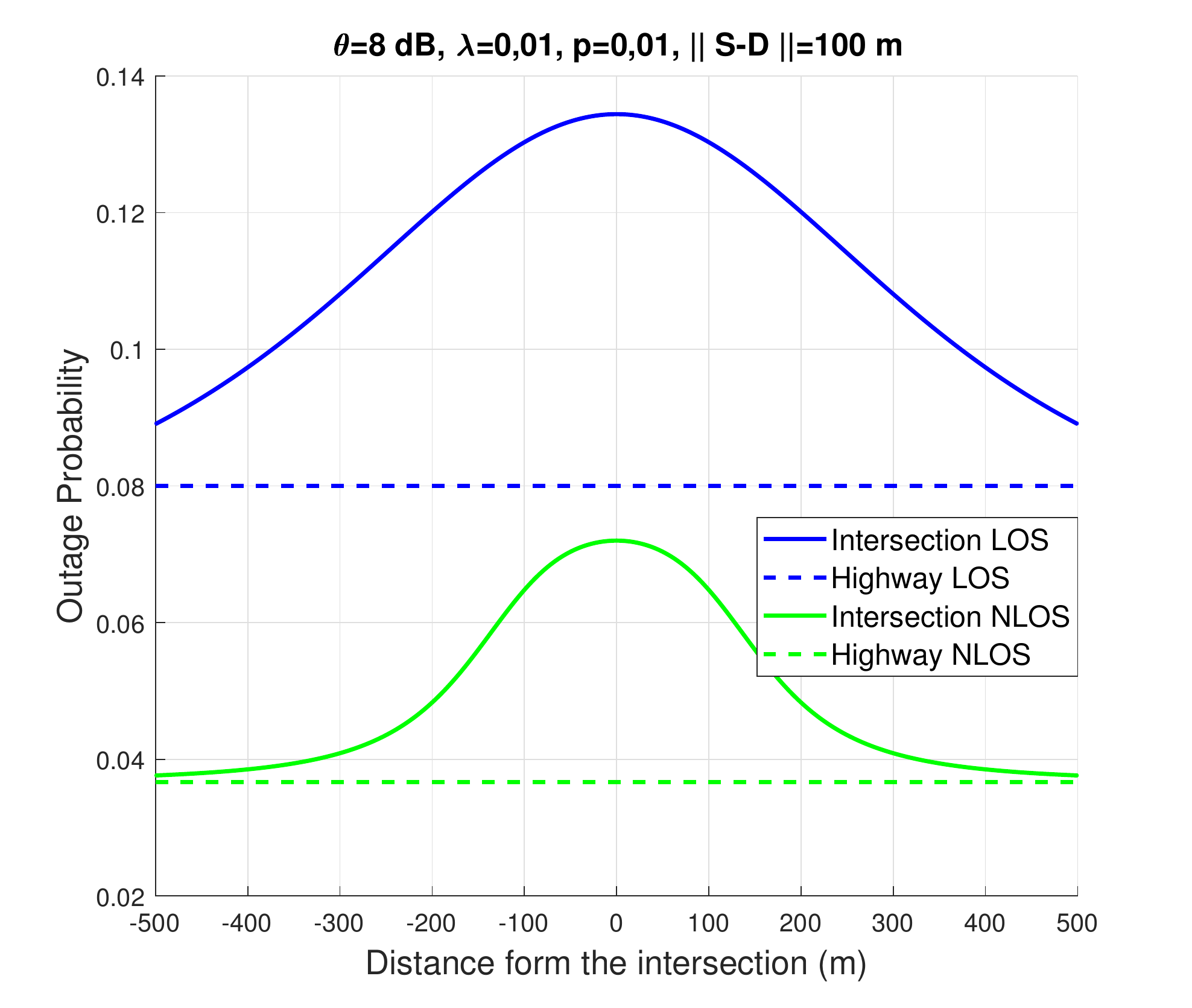}
        \caption{}
        \label{Fig3(a)}
    \end{subfigure}%
    ~ 
    \begin{subfigure}[b]{0.5\textwidth}
        \centering
        \includegraphics[height=8cm,width=9cm]{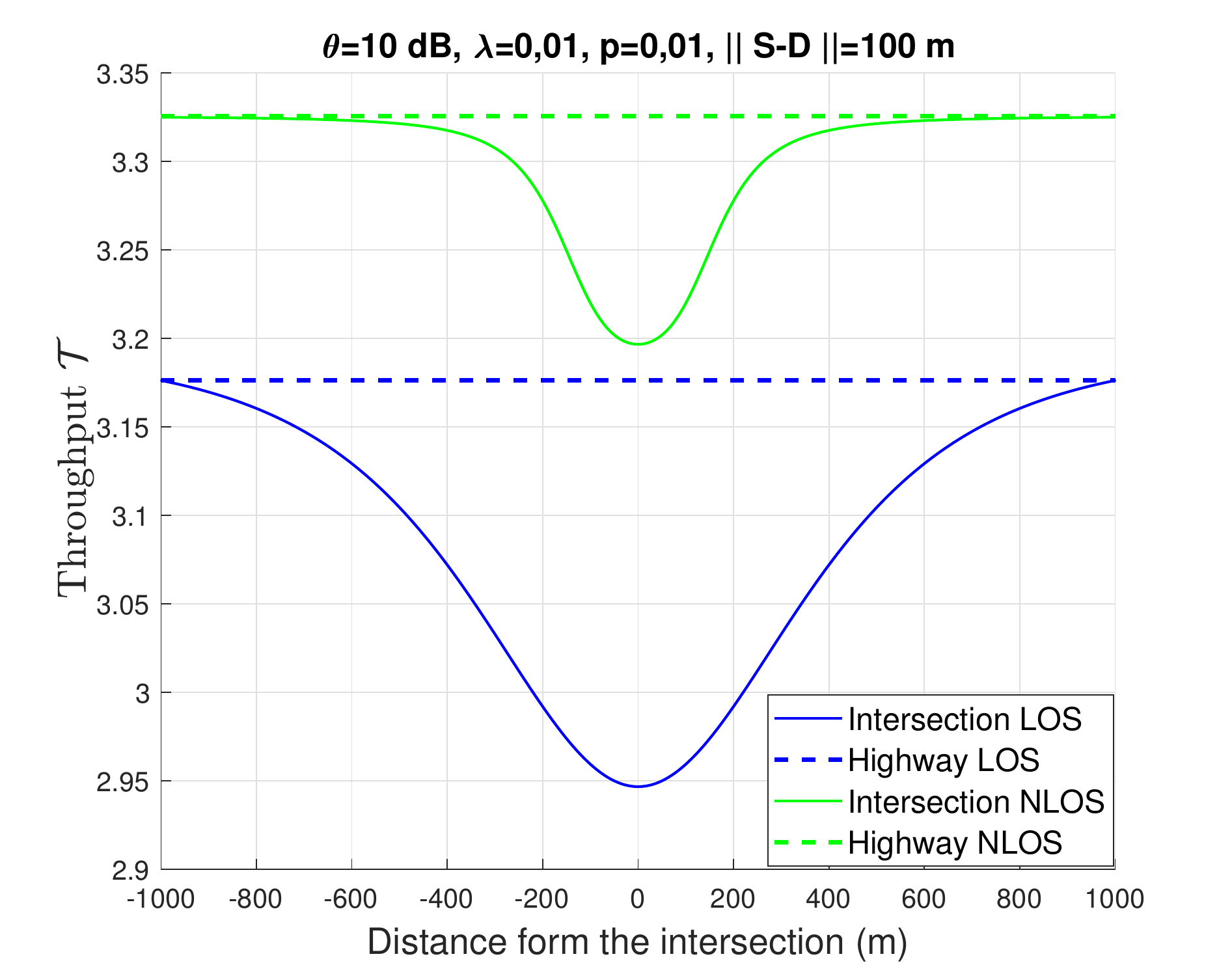}
        \caption{}
        \label{Fig3(b)}
    \end{subfigure}
    \caption{Performance of VCs as a function of the distance between the transmitting nodes and the interaction. (a) Outage probability for suburban environment (LOS), and urban environment (NLOS).(b)  throughput for suburban environment (LOS), and urban environment (NLOS).}
 \label{Fig3}
\end{figure*}

Fig.\ref{Fig3} shows the performance of VCs transmission when $S$ and $D$ move towards the intersection. We notice that, when $S$ and $D$ move toward the intersection, the performance decrease. This is because at intersection, all the interfering vehicles interfere at $D$, which is not the case when $D$ is far from the intersection. We also notice, as in Fig.\ref{Fig2}, that the LOS scenario has less performance that NLOS scenario. For instance, in Fig.\ref{Fig3(a)}, the throughput in NLOS scenario starts to decrease when $D$ is at 400 m from the intersection, whereas the throughput in LOS scenario starts to decrease when $D$ is at 1000 m from the intersection. For comparison purposes, we compared an intersection scenario with a highway scenario. We can see that the two scenarios offer the same performance when $D$ is far from the intersection. However, a significant increase in outage probability can be noticed when $D$ is at the intersection. For instance, the outage probability at intersection is 68\% higher than in highway for LOS scenario.
\begin{figure}[]
\centering
\includegraphics[height=8cm,width=9cm]{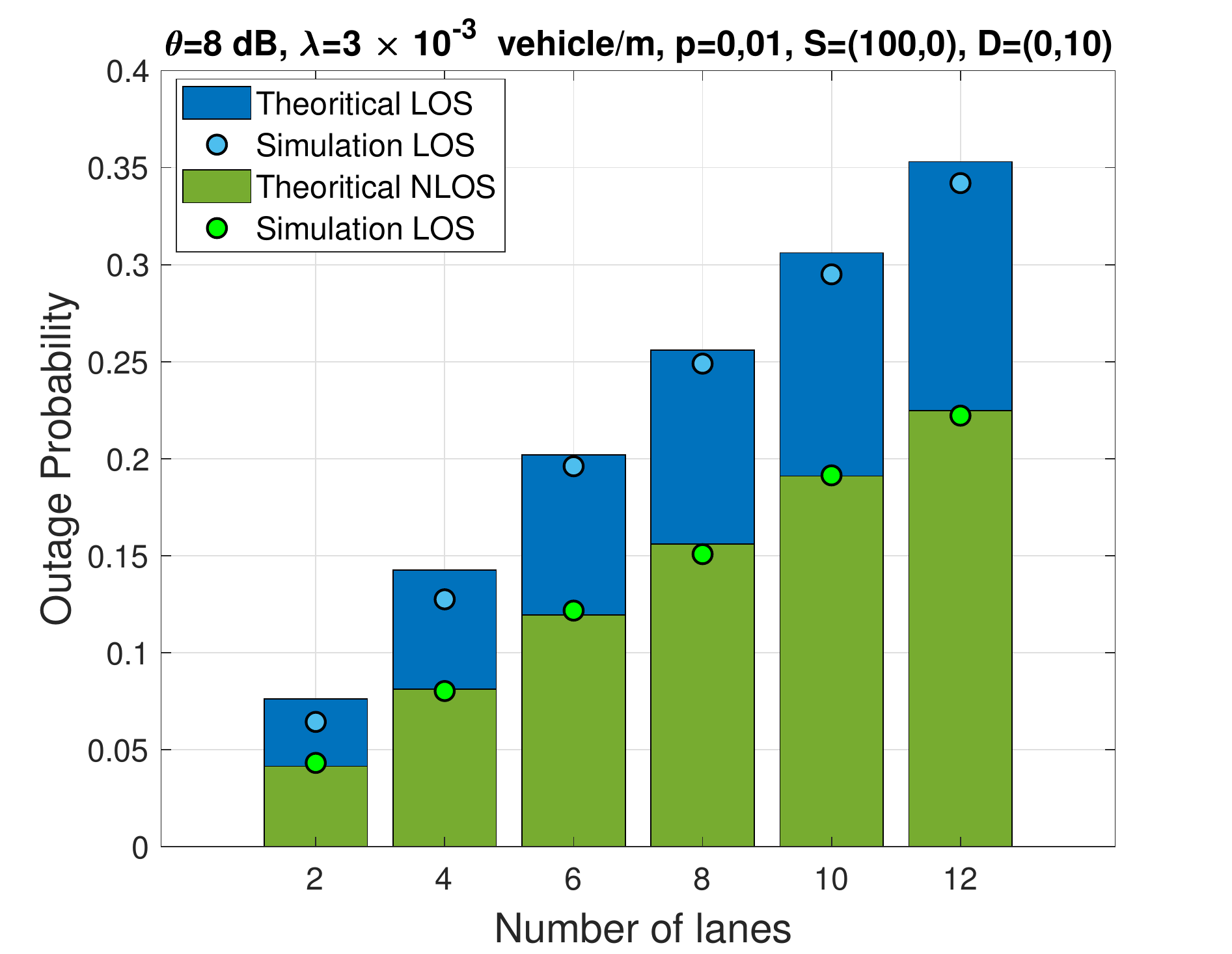}
\caption{Outage probability as a function of the number of lanes, for suburban environment (LOS), and urban environment (NLOS).}
\label{Fig4}
\end{figure}

Fig.\ref{Fig4} shows the performance in terms of outage probability when the number of lanes increases. We can see that as the number of lanes increases, the outage probability increases accordingly. 
This is because, when the number of lanes increases, the number of interfering vehicles increases, which increases the interference at $D$, hence increasing the outage probability. We also see that the outage probability increases linearly with the number of lanes. Therefore, the gap in performance between LOS and NLOS scenario increases as the number of lanes increases.
\section{Conclusion}
In this paper, we studied VCs at intersections, in the presence of interference over Nakagami-$m$ fading channels. We derived the outage probability expressions for specific channel conditions, and when the nodes are either on the road or outside the road. We compared the performance of the  suburban (LOS) environment and the urban (NLOS) environment, and showed that the urban environment  exhibits better performance than the suburban environment.
We also compared intersection scenarios with highway scenarios, and showed that the performance of intersection scenario are worst than highway as the destination move toward the intersection. Finally, we investigated a realistic scenario involving many lanes, and showed that the increase of lanes decreases the performance.

\bibliographystyle{ieeetr}
\bibliography{bibnoma}

\end{document}